\documentclass
[preprint,12pt,prd,eqsecnum,titlepage,showpacs,onecolumn,final,groupedaddress]{revtex4}%
\usepackage{amsfonts}
\usepackage{amssymb}
\usepackage{graphicx}
\usepackage{amsmath}%
\begin{document}
\title{Chameleon effect and the Pioneer anomaly}
\author{John D. Anderson\footnotemark\  \footnotetext{Retired}}
\affiliation{Jet Propulsion Laboratory, California Institute of Technology, Pasadena, CA
91109, USA}
\email{jdandy@earthlink.net}
\author{J.R. Morris}
\affiliation{Physics Department, Indiana University Northwest, 3400 Broadway, Gary, Indiana
46408, USA}
\email{jmorris@iun.edu}

\begin{abstract}
The possibility that the apparent anomalous acceleration of the Pioneer 10 and
11 spacecraft may be due, at least in part, to a chameleon field effect is
examined. A small spacecraft, with no thin shell, can have a more pronounced
anomalous acceleration than a large compact body, such as a planet, having a
thin shell. The chameleon effect seems to present a natural way to explain the
differences seen in deviations from pure Newtonian gravity for a spacecraft
and for a planet, and appears to be compatible with the basic features of the
Pioneer anomaly, including the appearance of a jerk term. However, estimates
of the size of the chameleon effect indicate that its contribution to the
anomalous acceleration is negligible. We conclude that any inverse-square
component in the anomalous acceleration is more likely caused by an unmodelled
reaction force from solar-radiation pressure, rather than a chameleon field effect.

\end{abstract}

\maketitle

\section{ \bigskip Introduction}

\ \ The Pioneer anomaly refers to an anomalous acceleration of the Pioneer 10
and 11 spacecraft that has been inferred for large heliocentric distances of
$\sim20-70$ AU, resulting from the presence of an anomalous Doppler
shift\cite{PA PRL98,Anderson02,PArev}. This small anomalous acceleration
$\vec{a}_{P}$, which is a deviation from the prediction of the Newtonian
acceleration $\vec{a}_{N}$, had previously been taken to have been an
essentially constant acceleration with a magnitude of $a_{P}=8.74\pm
1.33\times10^{-10}$ m/s$^{2}$. Recently, however, an analysis of more complete
data sets has supported the conclusion that the anomalous Pioneer acceleration
$\vec{a}_{P}$ actually decreases with time with a temporal decay rate of
magnitude $\dot{a}_{P}\approx1.7\times10^{-11}$ m/s$^{2}$/yr \cite{PA PRL
2011}. This anomalous acceleration is seen to act on both the Pioneer 10 and
11 spacecraft, and is directed sunward. In contrast, there appear to be no
such anomalous accelerations exhibited by planetary motions, disfavoring a
gravitational explanation, unless there is a modified theory of gravity where
small objects, such as spacecraft, are affected differently than planets.

\bigskip

\ \ There have been many attempts to explain the anomaly either due to mundane
causes or on the basis of new physics (see, e.g., \cite{PArev} and references
therein). However, it is possible that some combination of both of these gives
rise to the anomaly. Attention here is focused on the possibility that the
above features of the Pioneer anomaly may be explained, in a rather natural
way, by the chameleon effect \cite{KW PRD,KW PRL}. For an outward bound
spacecraft trajectory, the chameleonic acceleration decreases with distance
from the sun, and is therefore expected to give rise to a nonzero jerk term.

\bigskip

\ \ The basic aspects of the original Khoury-Weltman chameleon model are
briefly reviewed, along with the expression for the chameleonic acceleration
of a thick shelled spacecraft, due to the thin shelled sun. The basic features
of the Pioneer anomaly are presented and compared with those of the chameleon
model. Numerical estimates are made, including an estimate of the thin shell
factor for the sun, allowing a rough determination of the chameleonic
acceleration. It is concluded that for a chameleon-matter coupling constant of
order unity, $\beta\sim O(1)$, the chameleon acceleration is negligible in
comparison to the anomalous Pioneer acceleration, and the chameleonic jerk
term is negligible in comparison to that reported recently in ref.\cite{PA PRL
2011}. In addition, we simply apply solar system constraints on the PPN
parameter $\gamma$ obtained from the Cassini mission\cite{HF},\cite{Bertotti},
ignoring assumptions concerning the chameleon coupling to matter and estimates
of the sun's thin shell factor, and again find that the chameleonic
acceleration, along with the chameleonic jerk term, are negligible in
comparison to those reported for the Pioneer anomaly. We conclude that an
explanation of the Pioneer anomaly must likely lie elsewhere.

\section{An Inverse Square Component in Recent Measurements of the Pioneer
Anomaly}

In a recent paper Turyshev et al. \cite{PA PRL 2011} analyze archived radio
Doppler data extending back to February 14, 1979 for Pioneer 10 and January
12, 1980 for Pioneer 11. They produce a record of unmodelled radial
acceleration $a_{r}$ at a two-year sample interval for both spacecraft. We
plot these accelerations as a function of radial distance from the Sun in
Fig.~\ref{Fig1}. The plotted points can be fit with a simple inverse square
curve for each spacecraft, as shown by the two solid lines. When this inverse
square component is removed, the resulting accelerations are consistent with
the constant value reported previously \cite{Anderson02}. In addition, the
longer observation interval reveals a residual linear decrease in the
acceleration of (-0.024~$\pm$ 0.005)~$\times~10^{-10}$ m s$^{-2}$ per
astronomical unit (AU), much smaller than inferred by Turyshev et al. from the
$a_{r}$ data without the removal of the inverse-square curves. This inverse
square component is most likely a result of a mismodelling of solar radiation
pressure acting on the spacecraft. It is unlikely it results from
non-isotropic thermal emission from the spacecraft. Based on a model of the
spacecraft, including its power subsystem, Anderson et al. \cite{Anderson02}
conclude that the thermal contribution is (0.55 $\pm$ 0.55)~$\times~10^{-10}$
m s$^{-2}$ directed toward the Sun, and they account for it as a measurement
bias in their determination of the magnitude of the anomalous acceleration. It
is difficult to argue for anything more than a three-sigma thermal effect, or
a maximum contribution of 2.2~$\times~10^{-10}$ m s$^{-2}$, 25\% of the total
anomaly. The model used by Anderson et al. \cite{Anderson02} is given some
credence by its successful application to the Cassini spacecraft, where the
observed decrease in orbital energy is consistent with the model
\cite{Anderson10}. Even so, based on their own spacecraft model, Francisco et
al. \cite{Francisco2011} claim that the anomaly is 100\% thermal. In the
following we address the possibility that the observed inverse-square decrease
in the measured acceleration could indeed be a part of the anomaly. By means
of calculations based on the so-called Chameleon effect, we conclude that this
is unlikely.

\section{Chameleon effect}

\subsection{Equations of motion}

Basic features of the original chameleon model proposed by Khoury and Weltman
in refs.\cite{KW PRD,KW PRL} are summarized here, beginning with the Einstein
frame (EF) action%
\begin{equation}
S=\int d^{4}x\sqrt{g}\left\{  \frac{1}{2\kappa^{2}}\mathcal{R}[g_{\mu\nu
}]+\frac{1}{2}g^{\mu\nu}\partial_{\mu}\phi\partial_{\nu}\phi-V(\phi)\right\}
+S_{m}[A^{2}(\phi)g_{\mu\nu},\psi] \label{e1}%
\end{equation}

where $S_{m}$ is the matter portion of the action, containing the chameleon
scalar $\phi$ along with other fields represented collectively by $\psi$. A
metric with signature $(+,-,-,-)$ is used and $A(\phi)=e^{\beta\kappa\phi
}=\exp(\beta\phi/M_{0})$, with $\beta\ $a\ dimensionless\ coupling parameter,
assumed to be of order unity, $\kappa=\sqrt{8\pi G}=1/M_{0}$, where $M_{0}$ is
the reduced Planck mass. The EF metric $g_{\mu\nu}$ is related to the Jordan
frame (JF) metric $\tilde{g}_{\mu\nu}$ by $\tilde{g}_{\mu\nu}=A^{2}g_{\mu\nu}%
$. The matter portion of the action is%
\begin{equation}
S_{m}=\int d^{4}x\sqrt{\tilde{g}}\mathcal{\tilde{L}}_{m}(\tilde{g}_{\mu\nu
},\psi)=\int d^{4}x\sqrt{g}\mathcal{\mathcal{L}}_{m}[A^{2}(\phi)g_{\mu\nu
},\psi] \label{e2}%
\end{equation}
The action (\ref{e1}) gives rise to the equations of motion (EoM)%
\begin{equation}%
\begin{array}
[c]{ll}%
\mathcal{R}_{\mu\nu}-\frac{1}{2}g_{\mu\nu}\mathcal{R}=-\kappa^{2}%
\mathcal{T}_{\mu\nu}=-\kappa^{2}\left[  \mathcal{T}_{\mu\nu}^{\phi
}+\mathcal{T}_{\mu\nu}^{m}\right]  \smallskip & \\
\square\phi+\dfrac{\partial V}{\partial\phi}-\sigma=0; & \\
\dfrac{du^{\nu}}{ds}+\Gamma_{\alpha\beta}^{\nu}u^{\alpha}u^{\beta}-\dfrac
{1}{m}\partial_{\mu}m\left[  g^{\mu\nu}-u^{\mu}u^{\nu}\right]  =0 &
\end{array}
\label{e3}%
\end{equation}

where $\sigma\equiv\dfrac{\partial\mathcal{L}_{m}}{\partial\phi}$. For
nonrelativistic matter, $\sigma=-\beta\kappa\rho_{EF}=-\beta\kappa\bar{\rho
}A(\phi)$, where $\rho_{EF}=\mathcal{T}^{m}$ is the EF matter energy density
and $\bar{\rho}=\rho_{EF}A^{-1}(\phi)$ is a $\phi$ independent, conserved
energy density in the EF. The EoM for $\phi$ can therefore be written in the
form%
\begin{equation}
\square\phi+\frac{\partial V}{\partial\phi}+\beta\kappa\bar{\rho}A(\phi)=0
\label{e4}%
\end{equation}

and there is an effective potential $V_{eff}(\phi)=V(\phi)+\bar{\rho}%
A(\phi)=V(\phi)+\bar{\rho}e^{\beta\kappa\phi}$. Assuming a positive value of
$\beta$, and $V(\phi)$ to be a runaway potential, say of the form
$V=M^{5}/\phi$ (see, e.g., ref.\cite{KW PRD}), $V_{eff}$ develops a minimum at
a value of $\phi_{\min}$ which depends on the local energy density $\bar{\rho
}$. A large $\bar{\rho}$ results in a large chameleon mass $m_{\phi}%
^{2}=V_{eff}^{\prime\prime}(\phi_{\min})$, while a small $\bar{\rho}$ results
in a small mass $m_{\phi}$. \ 

\bigskip

\ \ Also note that there is an extra term involving $\partial_{\mu}(\ln
m)=\partial_{\mu}(\ln A(\phi))$ in the \textquotedblleft
geodesic\textquotedblright\ equation above. This arises from the fact that a
test mass having a constant value $m_{0}$ in the JF, corresponds to a $\phi$
dependent mass $m=m_{0}A(\phi)$ in the EF \cite{Dicke62}.

\bigskip

\ \ We will consider $\phi=\phi(r)$ to be a static weak field with a
dependence upon the radial distance from some source mass $M$, which generates
a Schwarzschild metrical gravity field, with $g_{00}=(1-\frac{r_{S}}%
{r})=(1-\frac{2GM}{r})$. The EoM in the Newtonian limit (weak field, static
limit, with nonrelativistic particle motion) yield%

\begin{equation}%
\begin{array}
[c]{l}%
\nabla^{2}h_{00}=2\kappa^{2}(\mathcal{T}_{00}-\frac{1}{2}\mathcal{T}_{\lambda
}^{\lambda})\smallskip\\
\nabla^{2}\phi-\dfrac{\partial V}{\partial\phi}-\beta\kappa\bar{\rho}%
A(\phi)=0\smallskip\\
\dfrac{d^{2}\vec{x}}{dt^{2}}=-\frac{1}{2}\nabla h_{00}-\nabla(\ln A)
\end{array}
\label{e5}%
\end{equation}
\bigskip

\textbf{Remarks:\ \ }(1) There are two contributions to the acceleration
$\vec{a}$ of a test mass, the metric or Newtonian part, $\vec{a}_{N}%
=-\nabla(\frac{1}{2}h_{00})$, and the scalar chameleon part, $\vec{a}%
_{c}=-\nabla(\ln A)=-\beta\kappa\nabla\phi$. So from the geodesic equation
above, $\vec{a}=\vec{a}_{N}+\vec{a}_{c}$.

\bigskip

(2) Assuming a chameleon-type of model as described by Khoury and Weltman,
where $V(\phi)$ is a decreasing function of $\phi$ and $A(\phi)$ is an
increasing function, the vacuum value $\phi_{c}$ gets shifted to smaller
values when $\bar{\rho}$ increases. The chameleon mass is given by $m_{\phi
}^{2}=V_{eff}^{\prime\prime}(\phi)=V^{\prime\prime}(\phi)+\beta^{2}\kappa
^{2}\bar{\rho}e^{\beta\kappa\phi}$. The mass $m_{\phi}$ is large where
$\bar{\rho}$ is large (and therefore the $\phi$ field is short ranged), but
where $\bar{\rho}$ becomes very small $m_{\phi}$ is very small, and the $\phi$
field becomes nearly massless and long ranged. Therefore, earth-based gravity
differs from deep space-based gravity.

\bigskip

(3) For a Schwarzschild metric the Newtonian gravitational field is
\begin{equation}
\vec{a}_{N}=-\tfrac{1}{2}\nabla g_{00}=-\tfrac{1}{2}\nabla(1-\frac{2GM}%
{r})=-\frac{GM}{r^{2}}\hat{r}=-\frac{r_{S}}{2r^{2}}\hat{r} \label{e6}%
\end{equation}

and the chameleon \textquotedblleft anomaly\textquotedblright\ is%
\begin{equation}
\vec{a}_{c}=-\beta\kappa\nabla\phi=-\beta\kappa\left(  \partial_{r}%
\phi\right)  \ \hat{r},\ \ \ \ \kappa=\sqrt{8\pi G}=1/M_{0} \label{e7}%
\end{equation}

(4) For a central mass $M$ located at $\vec{x}=0$, the matter part of
$\mathcal{T}_{\mu\nu}$ is $\mathcal{T}_{\mu\nu}^{m}=\delta_{\mu}^{0}%
\delta_{\nu}^{0}M\delta^{(3)}(\vec{x})$ and the chameleon field part is
$\mathcal{T}_{\mu\nu}^{\phi}=\partial_{\mu}\phi\partial_{\nu}\phi-\eta_{\mu
\nu}\left[  \frac{1}{2}\eta^{\alpha\beta}\partial_{\alpha}\phi\partial_{\beta
}\phi-V(\phi)\right]  $, with $\mathcal{T}_{\mu}^{\phi\mu}=\left(  \nabla
\phi\right)  ^{2}+4V(\phi)$. Then the $\phi$ contribution to the right hand
side of the first equation in (\ref{e5}) is%
\begin{equation}
\mathcal{T}_{00}^{\phi}-\frac{1}{2}\mathcal{T}_{\lambda}^{\phi\lambda}%
=-V(\phi)\text{\ \ \ for\ \ \ }\frac{r_{S}}{r}\ll1 \label{e8}%
\end{equation}
Inputing the Schwarzschild metric means that the stress-energy of the
chameleon field is assumed to have a negligible effect on the metric outside
of a source, like the sun, where $r\gg r_{S}=2GM$. This is expected to be the
case for small $\bar{\rho}$, large $\phi$, and small $V$.

\subsection{The chameleon field}

We adopt the chameleon model proposed by Khoury and Weltman in \cite{KW PRD,KW
PRL} and borrow their results. We consider a compact uniform spherical mass
$M_{c}$ with radius $R_{c}$. The exterior solution for the chameleon field
(see eq.(26) in ref.\cite{KW PRD}) is approximately given by%
\begin{equation}
\phi=-\frac{C}{r}e^{-m_{\infty}(r-R_{c})}+\phi_{\infty} \label{e9}%
\end{equation}

where $m_{\infty}$ is the chameleon mass outside of the object, $\phi_{\infty
}$ is the value of $\phi$ that minimizes $V_{eff}$ outside of the object, and
the density profile is given by%
\begin{equation}
\bar{\rho}=\left\{
\begin{array}
[c]{cc}%
\bar{\rho}_{c}, & r<R_{c}\\
\bar{\rho}_{\infty}, & r>R_{c}%
\end{array}
\right\}  \label{e10}%
\end{equation}

Inside the object, $V_{eff}$ is minimized by $\phi_{c}$ and the chameleon mass
is $m_{c}$. The constant $C$ takes a value%
\begin{equation}
C=\frac{\beta\kappa M_{c}}{4\pi}\left\{
\begin{array}
[c]{cc}%
\left(  \frac{3\Delta R_{c}}{R_{c}}\right)  , & \text{thin shell, }%
\frac{\Delta R}{R}\ll1\\
1, & \text{thick shell, }\frac{\Delta R}{R}>1
\end{array}
\right\}  \label{e11}%
\end{equation}

We define $\Delta_{c}=\Delta R_{c}/R_{c}$, which is given by (see eq.(16) of
\cite{KW PRD}),%
\begin{equation}
\Delta_{c}=\frac{\Delta R_{c}}{R_{c}}=\frac{\phi_{\infty}-\phi_{c}}{6\beta
M_{0}\Phi_{c}} \label{e12}%
\end{equation}

where $\Phi_{c}=M_{c}/8\pi M_{0}^{2}R_{c}=GM_{c}/R_{c}$ is the Newtonian
potential at the surface of the sphere, and $\phi\approx\phi_{c}$ well inside
the object, near the core, where the chameleon mass $m$ is large, $m_{c}\gg
m_{\infty}$. We then have%

\begin{equation}
\partial_{r}\phi=\left(  m_{\infty}+\frac{1}{r}\right)  C\left[
\frac{e^{-m_{\infty}(r-R_{c})}}{r}\right]  \label{e13}%
\end{equation}

As pointed out in \cite{KW PRD} and \cite{KW PRL}, a thin shelled object (like
a planet or a star) has a value of $C$, and hence the spatially varying part
of $\phi$, suppressed by a factor of $\Delta_{c}\ll1$ compared to a thick
shelled object (like a small satellite).

\subsection{Acceleration}

We define the radial component of acceleration by $\mathcal{A}=\hat{r}%
\cdot\vec{a}=a_{r}$. From (\ref{e6}), (\ref{e7}), and (\ref{e13}) we have the
Newtonian and chameleonic accelerations
\begin{subequations}
\label{e14}%
\begin{align}
\mathcal{A}_{N}  &  =-\frac{GM_{c}}{r^{2}}\label{e14a}\\
\mathcal{A}_{c}  &  =-\beta\kappa\partial_{r}\phi=-\beta\kappa C\left(
m_{\infty}+\frac{1}{r}\right)  \left[  \frac{e^{-m_{\infty}(r-R_{c})}}%
{r}\right]  \label{e14b}%
\end{align}

Both accelerations are directed radially inward ($\beta>0$), with
$a_{r}=\mathcal{A}_{N}+\mathcal{A}_{c}$, and the chameleonic acceleration acts
as an anomalous acceleration, i.e., a deviation from the Newtonian acceleration.

\bigskip

\ \ We now consider the case where $m_{\infty}r\ll1$, $m_{\infty}(r-R_{c}%
)\ll1$, and $r_{S}/r\ll1$, that is, for distances well outside a compact body
of mass $M_{c}$ and radius $R_{c}$. For a very small mass $m_{\infty}$ these
can be satisfied for distances $r\gg R_{c}$ so that $R_{c}\ll r\ll1/m_{\infty
}$. Assuming this to be the case, the chameleonic acceleration of a (thick
shelled) test mass is approximately%
\end{subequations}
\begin{equation}
\mathcal{A}_{c}\approx-\beta\kappa C\frac{1}{r^{2}} \label{e15}%
\end{equation}

Comparing this to the radial part of the Newtonian acceleration,%
\begin{equation}
\frac{\mathcal{A}_{c}}{\mathcal{A}_{N}}\approx2\beta^{2}\left\{
\begin{array}
[c]{cc}%
3\Delta_{c}, & \text{thin shell, }\Delta_{c}\ll1\\
1, & \text{thick shell, }\Delta_{c}>1
\end{array}
\right\}  \label{e16}%
\end{equation}

where (\ref{e11}) has been used and $\Delta_{c}$ is given by (\ref{e12}).
Therefore, for a large thin shelled source, like the sun, with $\Delta_{c}%
\ll1$, the chameleonic acceleration of a small thick shelled test mass is a
very small fraction of the Newtonian acceleration, with $\mathcal{A}%
_{c}\approx-6\beta^{2}\Delta_{c}GM_{c}/r^{2}$, with $R_{c}\ll r\ll1/m_{\infty
}$.

\bigskip

\ \ If the test mass is actually a thin shelled object, there is an additional
factor of $3\Delta$ for the test mass (see sec.VII A of \cite{KW PRD}), so
that the chameleonic acceleration of the thin shelled test mass due to a thin
shelled source is%
\begin{equation}
\frac{\mathcal{A}_{c}}{\mathcal{A}_{N}}\approx2\beta^{2}(3\Delta_{1}%
)(3\Delta_{2})=18\beta^{2}\Delta_{1}\Delta_{2} \label{e17}%
\end{equation}

This would describe the chameleon acceleration of a planet due to the sun, for
example, since both objects are thin shelled, and this acceleration is
suppressed by an additional $\Delta$ factor compared to that describing the
chameleon acceleration of a small thick shelled object, such as a small
satellite or spacecraft.

\bigskip

\ \ To summarize, for the condition $R_{c}\ll r\ll1/m_{\infty}$, there are
three cases:%
\begin{equation}%
\begin{array}
[c]{ll}%
\text{(i)} & \dfrac{\mathcal{A}_{c}}{\mathcal{A}_{N}}\approx2\beta^{2}%
(3\Delta_{S}\smallskip)=6\beta^{2}\Delta_{S}\\
\text{(ii)} & \dfrac{\mathcal{A}_{c}}{\mathcal{A}_{N}}\approx2\beta
^{2}(3\Delta_{1})(3\Delta_{2})=18\beta^{2}\Delta_{1}\smallskip\Delta_{2}\\
\text{(iii)} & \dfrac{\mathcal{A}_{c}}{\mathcal{A}_{N}}\approx2\beta
^{2}\smallskip
\end{array}
\label{e18}%
\end{equation}

where for the cases (i) - (iii) we have

(i) $S=$ source, thin shelled, $\Delta_{S}\ll1$; test particle is thick
shelled, $3\Delta\rightarrow1$

(ii) both source and test object are thin shelled; $\Delta_{1,2}\ll1$

(iii) both source and test particle are thick shelled, $(3\Delta_{1}%
)(3\Delta_{2})\rightarrow1$

\section{The Pioneer anomaly and the chameleon effect}

The Pioneer anomaly is associated with the observed deviations from predicted
Newtonian accelerations of the Pioneer 10 and 11 spacecraft after passing
about 20 AU from the sun, leaving the solar system. This anomalous
acceleration is very small, and often explained by unmodelled mundane causes,
but has some interesting features that seem compatible with the existence of a
chameleon effect. Some of these features, mentioned in the introduction, are
listed here.

\subsection{Pioneer anomaly features}

(1) There is a small apparent acceleration, that was previously assumed
constant with a magnitude of $a_{P}=8.74\pm1.33\times10^{-10}$ m/s$^{2}$, for
distances of $\sim20-70$ AU from the sun, with $a_{P}/a_{N}\ll1$. However, it
has recently been argued that the magnitude of $a_{P}$ has a (decreasing) time
dependence\cite{PA PRL 2011}. Specifically, a linear model with $a_{P}%
(t)=a_{P}(t_{0})+\dot{a}_{P}(t-t_{0})$ contains a jerk term $\dot{a}_{P}$,
with a reported value of $\dot{a}_{P}\approx-1.7\times10^{-11}$ m/s$^{2}$/yr
\cite{PA PRL 2011}.

(2) It seems to be directed inward toward the sun. (However \cite{PA PRL 2011}
report that the direction of the acceleration $\vec{a}_{P}$ remains
imprecisely determined, with no support for an inward direction toward the sun
over a direction toward the earth.)

(3) The same anomalous acceleration is seen for both spacecraft.

(4) Such anomalies are not observed in planetary motions, disfavoring a
gravitational explanation, unless a modified theory of gravity operates where
small objects, such as spacecraft, are affected differently than planets.

\subsection{Chameleon effect features}

Let us consider the simplistic interaction between a small, thick shelled
spacecraft and the massive, thin shelled sun.

\bigskip

\ \ (1) Case (i) of (\ref{e18}) gives a chameleon acceleration of%
\begin{equation}
\mathcal{A}_{c}\approx6\beta^{2}\Delta_{S}\mathcal{A}_{N}=-6\beta^{2}\left(
\frac{\Delta R_{S}}{R_{S}}\right)  \frac{GM_{S}}{r^{2}} \label{e19}%
\end{equation}

where $\Delta_{S}=(\Delta R_{S}/R_{S})\ll1$ for a thin shelled sun with radius
$R_{S}$. This satisfies $\mathcal{A}_{c}/\mathcal{A}_{N}\ll1$ for $\beta\sim
O(1)$ for a large thin shelled sun and a small thick shelled spacecraft. But
for $r=r(t)$ (\ref{e19}) indicates that a time dependence is present,
$\mathcal{A}=\mathcal{A}(t)$. The time rate of change of $|\mathcal{A}_{c}|$,
for a radial trajectory with velocity $v_{r}=dr/dt$, is%
\begin{equation}
\frac{d|\mathcal{A}_{c}|}{dt}=v_{r}\frac{d|\mathcal{A}_{c}|}{dr}\approx
-\frac{2}{r}v_{r}|\mathcal{A}_{c}(r)|,\text{\ \ or}\ \ \ \ \frac
{d(\ln|\mathcal{A}_{c}|)}{dt}\approx-\frac{2v_{r}}{r} \label{e20}%
\end{equation}

This is very small for a nonrelativistic speed $v_{r}\ll1$ and large distances
$r$, so we find $\mathcal{A}_{c}/\mathcal{A}_{N}\ll1$, with $|\mathcal{A}%
_{c}(t)|$ a slightly decreasing function of time for an increasing $r(t)$.

\ \ (2) The direction of $\mathcal{A}_{c}$ is radially inward, toward the sun.

\ \ (3) For two thick shelled spacecraft, the chameleon acceleration
$\mathcal{A}_{c}$ is the same for a given $r$, as seen in (\ref{e19}). For
very mildly varying $\mathcal{A}(r)$, the two Pioneer spacecraft should have
chameleon accelerations nearly the same, $\mathcal{A}_{c,10}\approx
\mathcal{A}_{c,11}$.

\ \ (4) The chameleon acceleration of a large thin shelled planet due to its
interaction with the thin shelled sun is suppressed by the planet's factor of
$3\Delta_{\text{planet}}=3\Delta R_{\text{planet}}/R_{\text{planet}}$, so that
from case (ii) of (\ref{e18}) we have%
\begin{equation}
\frac{\mathcal{A}_{c,\text{planet}}}{\mathcal{A}_{N}}\approx18\beta^{2}%
\Delta_{S}\Delta_{\text{planet}}=3\Delta_{\text{planet}}\frac{\mathcal{A}%
_{c,P}}{\mathcal{A}_{N}}\ll\frac{\mathcal{A}_{c,P}}{\mathcal{A}_{N}}
\label{e21}%
\end{equation}

where $\mathcal{A}_{c,P}$ is the chameleon acceleration of a Pioneer
spacecraft. The deviation from Newtonian acceleration is greatly suppressed
for a planet, and as pointed out in\cite{KW PRD,KW PRL} allows the chameleon
mechanism to easily pass all solar system tests of gravity.

\subsection{Numerical estimates}

Here, we make some approximate estimates based upon the original chameleon
model of Khoury and Weltman. The mass of the sun is $M_{S}=1.99\times10^{33}g$
and the distance of the earth from the sun is taken to be $r_{E}%
=1AU=1.5\times10^{13}cm$ which would give a Newtonian acceleration of the
earth toward the sun of $\mathcal{A}_{N,E}=-GM_{S}/r_{E}^{2}=-5.9\times
10^{-3}m/s^{2}$. Therefore the Newtonian acceleration of an object at a
distance $r$ from the sun is
\begin{equation}
\mathcal{A}_{N}=\mathcal{A}_{N,E}\left(  \frac{r_{E}}{r}\right)  ^{2}
\label{e22}%
\end{equation}
The Newtonian accelerations at distances of 20 AU and 70 AU are, respectively,%
\begin{equation}
\mathcal{A}_{N}^{20}=\mathcal{A}_{N,E}\left(  \frac{1}{20}\right)
^{2}=-1.5\times10^{-5}m/s^{2};\ \ \mathcal{A}_{N}^{70}=\mathcal{A}%
_{N,E}\left(  \frac{1}{70}\right)  ^{2}=-1.2\times10^{-6}m/s^{2} \label{e23}%
\end{equation}

The change in the magnitude of $\vec{a}_{N}$ between 20AU to 70AU is
\begin{equation}
\Delta|\vec{a}_{N}|=\Delta|\mathcal{A}_{N}|=|\mathcal{A}_{N}^{70}%
|-|\mathcal{A}_{N}^{20}|=-1.4\times10^{-5}m/s^{2} \label{e23a}%
\end{equation}

These results will be useful in estimates of space and time rates of change of
$\mathcal{A}_{c}$.

\ \ \textbf{Chameleon parameters:\ \ }From (\ref{e19}) we have a chameleonic
acceleration given by%

\begin{equation}
\frac{\mathcal{A}_{c}}{\mathcal{A}_{N}}\approx6\beta^{2}\Delta_{S}
\label{e24a}%
\end{equation}

provided that the Pioneer is thick shelled, i.e., $\Delta_{P}=\Delta
R_{P}/R_{P}>1$. We will assume this to be the case, so that an upper bound on
$\mathcal{A}_{c}$ is established with (\ref{e24a}). If the Pioneer were
actually thin shelled, with $\Delta_{P}\ll1$, there would be an additional
suppression factor of $3\Delta_{P}$ leading to a chameleon acceleration much
smaller than that of $6\beta^{2}\Delta_{S}$. We will also take $\beta\sim1$.

\bigskip

\ \ The upper bound on the thin shell factor $\Delta_{E}$ for the earth,
proposed by Khoury and Weltman in \cite{KW PRL} (see eq.(15)) is given as%
\begin{equation}
\Delta_{E}<10^{-7} \label{e32}%
\end{equation}
We can use this in our estimate for the shell factor $\Delta_{S}$ for the sun
that appears in (\ref{e24a}). The shell factor $\Delta_{S}$ for the sun, and
the shell factor $\Delta_{E}$ for the earth are taken to be
\begin{equation}
\Delta_{S}=\frac{\Delta R_{S}}{R_{S}}=\frac{\phi_{G}-\phi_{S}}{6\beta
M_{0}\Phi_{S}},\ \ \ \ \ \Delta_{E}=\frac{\Delta R_{E}}{R_{E}}=\frac{\phi
_{G}-\phi_{E}}{6\beta M_{0}\Phi_{E}} \label{2}%
\end{equation}

where $\phi_{S(E)}$ is the value of $\phi_{c}$ inside the sun (earth),
$\phi_{G}$ is the value of $\phi_{\infty}$ in our galaxy, and $\Phi_{S(E)}$ is
the Newtonian potential at the surface of the sun (earth), $\Phi=GM/R$. \ Now,
take $\rho_{E}\sim5.7$ g/cm$^{3}$, $\rho_{S}\sim1.4$ g/cm$^{3}\sim\frac{1}%
{4}\rho_{E}$; we take these to be roughly equal for simplicity, $\rho_{S}%
\sim\rho_{E}$, and since $\phi_{c}$ is determined by the density $\bar{\rho}$,
we therefore take $\phi_{S}\approx\phi_{E}$. (For a high density contrast,
$\bar{\rho}_{c}\gg\bar{\rho}_{\infty}$, we have $\phi_{c}\ll\phi_{\infty}$, so
that $\phi_{G}-\phi_{S(E)}\approx\phi_{G}$, and consequently $\Delta
_{S}/\Delta_{E}\approx\Phi_{E}/\Phi_{S}$.) For the Newtonian potentials,%
\begin{equation}
\frac{\Phi_{S}}{\Phi_{E}}=\frac{M_{S}}{M_{E}}\frac{R_{E}}{R_{S}}\approx
3\times10^{3};\ \ \ \ \Phi_{S}\approx3\times10^{3}\Phi_{E} \label{3}%
\end{equation}

From (\ref{2})%
\begin{equation}
\frac{\Delta_{S}}{\Delta_{E}}\sim\frac{\Phi_{E}}{\Phi_{S}}\approx
3\times10^{-4}\sim10^{-4};\ \ \ \ \ \Delta_{S}\sim10^{-4}\Delta_{E} \label{4}%
\end{equation}

\ \ Using $\Delta_{E}\sim10^{4}\Delta_{S}$, (\ref{4}) and (\ref{e32}) give
\begin{equation}
\Delta_{S}<10^{-11} \label{6}%
\end{equation}

\bigskip

\ \textbf{Chameleon acceleration:}\ \ From (\ref{e24a}) and (\ref{6})%
\begin{equation}
\frac{\mathcal{A}_{c}}{\mathcal{A}_{N}}\approx6\beta^{2}\Delta_{S}%
\lesssim6\beta^{2}\times10^{-11} \label{7}%
\end{equation}

The average contribution to the Pioneer anomalous acceleration would be
roughly%
\begin{equation}
\frac{\left\langle |\mathcal{A}_{c}|\right\rangle }{a_{P}}\approx6\beta
^{2}\Delta_{S}\frac{\left\langle |\mathcal{A}_{N}|\right\rangle }{a_{P}}
\label{8}%
\end{equation}

We can estimate a spatial average of the Newtonian acceleration,%
\begin{equation}
\left\langle |\mathcal{A}_{N}|\right\rangle =\left\langle a_{N}\right\rangle
=\frac{1}{\Delta r}\int_{r_{1}}^{r_{2}}\frac{GM_{S}}{r^{2}}dr=\frac{GM_{S}%
}{\Delta r}\frac{\Delta r}{r_{1}r_{2}}=\frac{GM_{S}}{r_{1}r_{2}} \label{9}%
\end{equation}

and taking $r_{1}=20r_{E}=20$AU and $r_{2}=70r_{E}=70$AU, we get%
\begin{equation}
\left\langle |\mathcal{A}_{N}|\right\rangle =\left\langle a_{N}\right\rangle
\approx4.2\times10^{-6}\ \text{m/s}^{2} \label{10}%
\end{equation}

Eqs. (\ref{6}) and (\ref{8}) then give%
\begin{equation}
\frac{\left\langle |\mathcal{A}_{c}|\right\rangle }{a_{P}}\sim6\beta^{2}%
\Delta_{S}\left[  \frac{4.2\times10^{-6}\text{ m/s}^{2}}{9\times10^{-10}\text{
m/s}^{2}}\right]  =\beta^{2}\Delta_{S}(2.8\times10^{4})<(2.8\times
10^{-7})\beta^{2} \label{10a}%
\end{equation}

\ \ Taking $a_{P}\sim10^{-9}$ m/s$^{2}$ and $\beta\approx1$, from (\ref{10a})
it appears that a chameleon acceleration, if it existed, would be undetectably
small, with an average value estimated as%
\begin{equation}
\left\langle |\mathcal{A}_{c}|\right\rangle \lesssim10^{-16}\text{ m/s}^{2}
\label{11}%
\end{equation}

If the Pioneer spacecraft were actually thin shelled, there would be an
additional suppression factor of $3\Delta_{P}\ll1$ according to case (ii) of
(\ref{e18}), reducing the chameleonic acceleration even further, so that
(\ref{11}) serves as an upper bound on $\left\langle |\mathcal{A}%
_{c}|\right\rangle $.

\bigskip

\textbf{Spatial and temporal variation:}\ \ We can use a linear model to
estimate an average space rate of change $\Delta|\mathcal{A}_{c}|/\Delta r$,
for a change in distance of 50 AU from $r_{1}=$ 20 AU to $r_{2}=$ 70 AU and
using (\ref{e23a}):%
\begin{equation}
\frac{\Delta|\mathcal{A}_{c}|}{\Delta r}\approx6\beta^{2}\Delta_{S}%
\frac{\Delta|\mathcal{A}_{N}|}{\Delta r}\sim6\beta^{2}\Delta_{S}\left[
\frac{|\mathcal{A}_{N}^{70}|-|\mathcal{A}_{N}^{20}|}{50\ \text{AU}}\right]
\sim6\beta^{2}\Delta_{S}(-2.7\times10^{-7}\text{m/s}^{2}/\text{AU}) \label{12}%
\end{equation}

Therefore (\ref{7}) gives%
\begin{equation}
\Big|\frac{\Delta|\mathcal{A}_{c}|}{\Delta r}\Big|<(6\beta^{2}\times
10^{-11})(2.7\times10^{-7}\text{m/s}^{2}/\text{AU})\sim10^{-17}\text{m/s}%
^{2}/\text{AU} \label{13}%
\end{equation}

So the estimate of
\begin{equation}
\Big|\frac{\Delta|\mathcal{A}_{c}|}{\Delta r}\Big|<10^{-17}\text{m/s}%
^{2}/\text{AU} \label{14}%
\end{equation}

is negligible in size in comparison to an estimate of
\begin{equation}
\Big|\frac{\Delta a_{P}}{\Delta r}\Big|\sim|\dot{a}_{P}|\frac{\Delta t}{\Delta
r}\sim(.17\times10^{-10}\text{m/s}^{2}/\text{yr})\frac{30\text{yr}%
}{50\text{AU}}\sim10^{-11}\text{m/s}^{2}/\text{AU} \label{15}%
\end{equation}
obtained using the jerk term $\dot{a}_{P}$ in ref.\cite{PA PRL 2011}.

\bigskip

\ \ A chameleonic jerk term $\dot{a}_{c}$ can be estimated using $\dot{a}%
_{c}=-\mathcal{\dot{A}}\approx\Delta|\mathcal{A}|/\Delta t$, with
$\Delta|\mathcal{A}|\approx|\mathcal{A}_{70}|-|\mathcal{A}_{20}|$ and $\Delta
t\sim30$ yr; this will be a negative number since $r$ decreases with time for
an outward bound trajectory and $|\mathcal{A}|\propto a_{N}\propto1/r^{2}$. We
have%
\begin{equation}
|\dot{a}_{c}|\sim6\beta^{2}\Delta_{S}\Big|\dfrac{\Delta|a_{N}|}{\Delta
t}\Big|\lesssim(6\beta^{2}\times10^{-11})\dfrac{\left(  1.4\times
10^{-5}\text{m/s}^{2}\right)  }{30\text{yr}}\sim2.7\times10^{-17}%
\text{m/s}^{2}\text{/yr} \label{16}%
\end{equation}

The value of the Pioneer jerk term (using the linear model) reported in
ref.\cite{PA PRL 2011} is $|\dot{a}_{P}|=.17\times10^{-10}$m/s$^{2}$/yr, so
that $\dot{a}_{c}/\dot{a}_{P}\lesssim10^{-6}$. From (\ref{10a}), (\ref{11}),
and (\ref{14})-(\ref{16}), we conclude that within the context of the original
Khoury-Weltman model, the chameleon effect, if it exists, is too small to
account for the anomalous Pioneer acceleration or its spatial or temporal rate
of change reported in \cite{PA PRL 2011}.

\section{Solar system constraints}

In the previous section we have assumed, as in the original chameleon model of
Khoury and Weltman, that $\beta\approx1$ and we have used their results to
obtain the estimate for the thin shell factor for the sun $\Delta_{S}%
<10^{-11}$. We see (e.g., from Eq.(\ref{10a})) that the chameleonic
contribution to the Pioneer anomaly is controlled by the factor $\beta
^{2}\Delta_{S}$. We now use this result, but relax our assumption that
$\beta\sim1$ and abandon our estimate of $\Delta_{S}$, and instead, obtain a
fix on the factor $\beta^{2}\Delta_{S}$ by using the results obtained in the
recent analysis by Hees and Fuzfa \cite{HF}, wherein an upper limit of this
factor can be obtained from the PPN parameter $\gamma$ obtained from solar
system contraints of the Cassini mission\cite{Bertotti}:%
\begin{equation}
\gamma-1=(2.1\pm2.3)\times10^{-5} \label{17}%
\end{equation}

\ \ Hees and Fuzfa (HF) use slightly different notations for the scalar field
and chameleon parameters, but we can readily build a simple translation
dictionary by noting that HF write the action in a form (using our metric
signature)%
\begin{equation}
S=\int d^{4}x\sqrt{g}\left\{  \frac{m_{P}^{2}}{16\pi}\mathcal{R}[g_{\mu\nu
}]+\frac{1}{2}m_{P}^{2}g^{\mu\nu}\partial_{\mu}\hat{\phi}\partial_{\nu}%
\hat{\phi}-\hat{V}(\hat{\phi})\right\}  +S_{m}\left[  \hat{A}^{2}(\hat{\phi
})g_{\mu\nu},\psi\right]  \label{18}%
\end{equation}

where the hat notation denotes the fields and functions used by HF,
$m_{P}=1/\sqrt{G}$, $\hat{A}(\hat{\phi})=e^{k\hat{\phi}}$, and the JF metric
$\tilde{g}_{\mu\nu}$ and EF metric $g_{\mu\nu}$ are related by $\tilde{g}%
_{\mu\nu}=\hat{A}(\hat{\phi})g_{\mu\nu}$. Comparison with (\ref{e1}) then
shows that%
\begin{equation}
\hat{\phi}=\phi/m_{P},\ \ \hat{A}(\hat{\phi})=e^{k\hat{\phi}}=A(\phi
)=e^{\beta\kappa\phi},\ \ k=\sqrt{8\pi}\beta,\ \ \ \hat{V}(\hat{\phi})=V(\phi)
\label{19}%
\end{equation}

\ \ It should be noted, however, that\cite{HF} use a different form of the
effective potential, as they choose to use the Jordan frame energy density
$\tilde{\rho}$ as a constant, rather than the conventionally chosen density
$\bar{\rho}=\rho_{EF}A^{-1}(\phi)$, which is a $\phi$ independent quantity in
the Einstein frame representation of the theory\cite{KW PRD,KW PRL}. Thus, the
HF effective potential is written as%
\begin{equation}
\hat{V}_{eff}(\hat{\phi})=\hat{V}(\hat{\phi})+\frac{1}{4}\tilde{\rho}%
e^{4k\hat{\phi}} \label{20}%
\end{equation}

instead of our conventionally chosen effective potential (see Eq.(\ref{e4}))%
\begin{equation}
V_{eff}(\hat{\phi})=\hat{V}(\hat{\phi})+\bar{\rho}e^{k\hat{\phi}}=V(\phi
)+\bar{\rho}e^{\beta\kappa\phi} \label{21}%
\end{equation}

This difference can be largely ignored, however, as it does not qualitatively
change the results obtained\cite{HF}. More specifically, we borrow the result
from \cite{HF} that $k\hat{\phi}_{\infty}\lesssim2\times10^{-12}\ll1$, where
$\hat{\phi}_{\infty}$ minimizes the effective potential at $r=\infty$, so that
for $\tilde{\rho}\approx\bar{\rho}$ and fields of interest where $\hat{\phi
}\leq\hat{\phi}_{\infty},$ we have the ratio%
\begin{equation}
\frac{\frac{1}{4}\tilde{\rho}\hat{A}^{4}}{\bar{\rho}\hat{A}}=\frac{1}{4}%
\frac{\tilde{\rho}}{\bar{\rho}}e^{3k\hat{\phi}}\approx O(1) \label{22}%
\end{equation}

showing that we have reasonable confidence in using our effective potential
along with an application of the results obtained in \cite{HF}.

\subsection{Cassini bounds}

\ \ Hees and Fuzfa obtain the result relating the effective coupling constant
$k_{eff}$, the thin shell factor $\epsilon=\Delta_{S}$ for the sun, and the
PPN parameter, $\gamma$%
\begin{equation}
(\gamma-1)=-\frac{2kk_{eff}}{4\pi+kk_{eff}}\approx-6\frac{\epsilon k^{2}}%
{4\pi+3\epsilon k^{2}} \label{23}%
\end{equation}

In order for the chameleon mechanism to account for a nonzero value of
$(\gamma-1)$, we see that $(\gamma-1)$ must be negative, so that from
(\ref{17})%
\begin{equation}
|\gamma-1|\leq|\gamma-1|_{\max}=2\times10^{-6} \label{24}%
\end{equation}

Inverting (\ref{23}) to obtain $\epsilon k^{2}$, we have%
\begin{equation}
(\epsilon k^{2})_{\max}\approx\frac{4\pi}{3}\left(  \frac{|\gamma-1|_{\max}%
}{2-|\gamma-1|_{\max}}\right)  \approx4.2\times10^{-6} \label{25}%
\end{equation}

In terms of our original KW parameters $\beta$ and $\Delta_{S}$, this
translates into%
\begin{equation}
\beta^{2}\Delta_{S}\lesssim3.3\times10^{-7} \label{26}%
\end{equation}

We note that this is in accord with our previous estimates based upon
$\beta\approx1$ and $\Delta_{S}<10^{-11}$.

\subsection{Estimates based upon the Cassini bounds}

\ \ \textbf{Chameleonic acceleration:}\ \ We can now simply use (\ref{26})
without any assumptions for the values of $\beta$ and $\Delta_{S}$ to obtain
estimates of the maximum chameleonic contribution to the Pioneer anomaly. For
example, using (\ref{26}) in (\ref{10a}) yields%
\begin{equation}
\frac{\left\langle |\mathcal{A}_{c}|\right\rangle }{a_{P}}\lesssim
5.5\times10^{-2} \label{27}%
\end{equation}

indicating that a chameleonic acceleration could account for no more than
$5.5\%$ of the Pioneer acceleration.

\bigskip

\ \ \textbf{Spatial and temporal variation:}\ \ In a similar manner, referring
back to Eqs.(\ref{12})-(\ref{16}), the application of (\ref{26}) gives a
spatial variation%
\begin{equation}
\Big|\frac{\Delta|\mathcal{A}_{c}|}{\Delta r}\Big|\lesssim5.4\times
10^{-13}\text{ m/s}^{2}\text{/AU} \label{28}%
\end{equation}

and%

\begin{equation}
\Big|\frac{|\Delta\mathcal{A}_{c}|/\Delta r}{|\Delta a_{P}|/\Delta
r}\Big|\lesssim5.4\times10^{-2} \label{29}%
\end{equation}

and a time variation (jerk term)%
\begin{equation}
|\dot{a}_{c}|\lesssim9\times10^{-13}\text{ m/s}^{2}\text{/yr} \label{30}%
\end{equation}

with%
\begin{equation}
\frac{\dot{a}_{c}}{\dot{a}_{P}}\lesssim5.3\times10^{-2} \label{31}%
\end{equation}

Again, apparently the chameleonic jerk term is no more than about $5.3\%$ of
the reported Pioneer jerk term.

\section{Summary}

\ \ The chameleon model proposed in \cite{KW PRD,KW PRL} has basic features
that seem to be compatible, in a natural way, with the prominent features
exhibited by the Pioneer anomaly. A small, thick shelled spacecraft can have a
much more pronounced deviation from a Newtonian acceleration than can a large,
massive, thin shelled planet. Therefore, the anomaly seen by the Pioneer 10
and 11 spacecraft does not become manifest in any anomalous planetary motions.

\bigskip

\ \ Furthermore, the chameleon effect produces an acceleration which is small
in comparison to the Newtonian acceleration if the spacecraft is thick
shelled, and this acceleration is directed sunward, i.e., toward the
gravitational source. The chameleonic acceleration is found to have a
$1/r^{2}$ dependence, so that for an outward bound journey the chameleon
\textquotedblleft anomaly\textquotedblright\ decreases in magnitude.

\bigskip

\ \ We have estimated the chameleonic acceleration and its spatial and
temporal rates of change, and conclude that the chameleon effect can not
account for the Pioneer anomalous acceleration or jerk term recently reported
by \cite{PA PRL 2011}. Specifically, using the original Khoury-Weltman
chameleon model and results, we find that $\left\langle |\mathcal{A}%
_{c}|\right\rangle /a_{P}\lesssim10^{-7}$, $\frac{\Delta|\mathcal{A}%
_{c}|/\Delta r}{\Delta a_{P}/\Delta r}\lesssim10^{-6}$, and $\dot{a}_{c}%
/\dot{a}_{P}\lesssim10^{-6}$.

\bigskip

\ \ However, more general considerations simply based upon solar system
constraints (specifically the constraints from the Cassini bounds on the PPN
parameter $\gamma$), lead to maximum contributions $\left\langle
|\mathcal{A}_{c}|\right\rangle /a_{P}\lesssim5.5\times10^{-2}$, $\frac
{\Delta|\mathcal{A}_{c}|/\Delta r}{\Delta a_{P}/\Delta r}\lesssim
5.4\times10^{-2}$, and $\dot{a}_{c}/\dot{a}_{P}\lesssim5.3\times10^{-2}$. We
conclude that solar system constraints allow possible chameleonic effects to
account for no more than a few percent of those that are observed. We suspect
that an inverse square component seen in the anomalous acceleration is more
likely due to an unmodelled reaction force from solar-radiation pressure,
rather than a chameleon field effect.

\smallskip\ \ \medskip\textbf{Acknowledgement:} We wish to thank an anonymous
referee for useful comments.

\newpage

\section*{Figures}

%

\begin{center}
\includegraphics[
height=4.4483in,
width=6.2914in
]%
{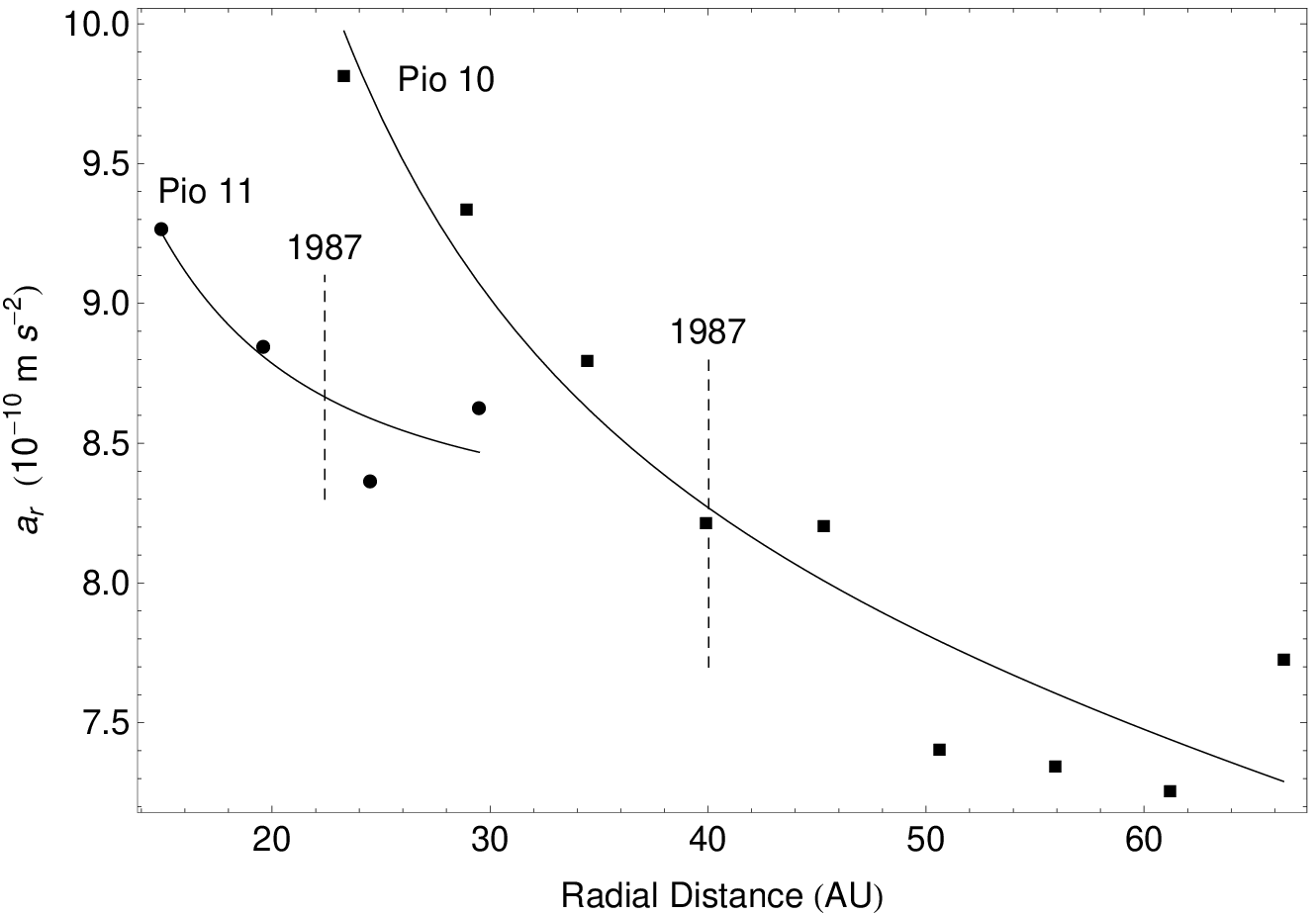}%
\end{center}

\begin{figure}[th]
\begin{center}
\noindent
\end{center}
\caption{Measured values of unmodelled radial acceleration $\mathrm{a_{r}}$
(in units of $\mathrm{10^{-10}~m~s^{-2}}$) according to Turyshev et al.
\cite{PA PRL 2011}, but plotted as a function of radial distance r in
astronomical units (AU) rather than time. The two fitting curves are given by
the function $\mathrm{k_{0}+k_{1}r+k_{2}/r^{2}}$, where $\mathrm{k_{1}}$ is
set to zero for Pioneer 11. The two dashed lines indicate the radii at the
beginning of 1987. No data prior to 1987 were used in obtaining the anomalous
acceleration of $\mathrm{\left(  8.74\pm1.33\right)  \times10^{-10}~m~s^{-2}}$
reported by Anderson et al. \cite{Anderson02}, although after subtraction of
the inverse-square component $\mathrm{k_{2}/r^{2}}$, and with a reported
measurement bias of $\mathrm{0.90\times10^{-10}~m~s^{-2}}$ added in
\cite{Anderson02}, the resulting accelerations are well within the standard
error of the 2002 result. }%
\label{Fig1}%
\end{figure}

\end{document}